\documentclass[article]{emulateapj}
\usepackage{amsmath}
\usepackage{lineno}

\slugcomment{submitted; \today}

\newcommand{\G}{\Gamma}

\newcommand{\e}{\epsilon}

\newcommand{\g}{\gamma}

\newcommand{\ep}{\epsilon^\prime}

\newcommand{\psim}{\lower.5ex\hbox{$\; \buildrel \propto \over\sim \;$}}
\newcommand{\lbar}{\lower.0ex\hbox{$\; \buildrel{\lower0.0ex \hbox{-}} \over\lambda  \;$}}

\newcommand{\m}{\mathrm{m}}
\newcommand{\cm}{\mathrm{cm}}
\newcommand{\km}{\mathrm{km}}

\newcommand{\eV}{\mathrm{eV}}

\newcommand{\GeV}{\mathrm{GeV}}
\newcommand{\TeV}{\mathrm{TeV}}
\newcommand{\s}{\mathrm{s}}

\newcommand{\years}{\mathrm{years}}

\newcommand{\Mpc}{\mathrm{Mpc}}

\newcommand{\ph}{\mathrm{ph}}

\shorttitle{Lorentz Invariance Violation}
\shortauthors{Finke \& Razzaque}

\begin{document}

\title{Possible Evidence for Lorentz Invariance Violation in Gamma-ray Burst 221009A}

\author{Justin D.\ Finke}

\affil{U.S.\ Naval Research Laboratory, Code 7653, 4555 Overlook Ave.\ SW,
        Washington, DC,
        20375-5352, USA; \\
        justin.finke@nrl.navy.mil}

\author{Soebur Razzaque$^\dagger$}\altaffiliation[$^\dagger$Also at the ]{Department of Physics, The George Washington University, Washington, DC 20052, USA; and National Institute for Theoretical and Computational Sciences (NITheCS), South Africa}
\affil{Centre for Astro-Particle Physics (CAPP) and Department of Physics, \\
University of Johannesburg, P.O. Box 524, Auckland Park 2006, South Africa; \\ srazzaque@uj.ac.za \\}

\begin{abstract}

The preliminary detections of the gamma-ray burst 221009A up to 18 TeV by LHAASO and up to 251 TeV by Carpet 2 have been reported through Astronomer's Telegrams and Gamma-ray Coordination Network circulars.  Since this burst is at redshift $z=0.1505$, these photons may at first seem to have a low probability to avoid pair production off of background radiation fields and survive to reach detectors on Earth.  By extrapolating the reported $0.1-1.0$\ GeV LAT spectrum from this burst to higher energies and using this to limit the intrinsic spectrum of the burst, we show that the survival of the 18 TeV photon detected by LHAASO is not unlikely with many recent extragalactic background light models, although the detection of a 251 TeV event is still very unlikely.  This can be resolved if Lorentz invariance is violated at an energy scale $E_{\rm QG}\la 49 E_{\rm Planck}$\ in the linear ($n=1$) case, and $E_{\rm QG}\la 10^{-6}E_{\rm Planck}$\ in the quadratic ($n=2$) case (95\% confidence limits), where $E_{\rm Planck}$ is the Planck energy.  This could potentially be the first evidence for subluminal Lorentz invariance violation.

\end{abstract}



\section{Introduction}
\label{intro}

The gamma-ray burst (GRB) 221009A  (also known as Swift J1913.1+1946) was detected by the {\em Swift}-Burst Alert Telescope  \citep[BAT;][]{kennea22} and  the {\em Fermi} Gamma-ray burst Monitor \citep[GBM;][]{veres22} as the brightest GRB ever detected.  It was also detected by the {\em Fermi} Large Area Telescope \citep[LAT;][]{bissaldi22,pillera22}.
At a redshift of $z=0.1505$ \citep{deugarte22,izzo22,castro22} it is also one of the closest long-duration GRBs.

Perhaps most surprising is the possible detection of photons at $E>10\ \TeV$ from this burst.   In the 2000 s after the start of the burst ($T_0$), it was detected by the Large High Altitude Air Shower Observatory (LHAASO) with its WCDA and KM2A detectors, and the latter detected photons from GRB~221009A with energies up to 18 TeV \citep{huang22}. At $T_0 + 4536\ \s$, there was a report of an
astonishing 251 TeV photon detected from this burst by the Carpet 2 detector which has an  estimated probability of $1.2\times10^{-4}$ (corresponding to $3.8\sigma$; pre-trial) of being a background event \citep{dzhappuev22}.    There is a nearby HAWC source detected up to 140 TeV with a position consistent with both the reported LHAASO and Carpet 2 detection \citep{fraija22} that is probably Galactic.  This could be the source of the LHAASO detection, but it is unlikely to be the origin of the Carpet 2 detection; see Section \ref{hawcsection} below. 

Detection of these VHE photons from GRB~221009A is interesting for a number of reasons.  They may be difficult to explain with synchrotron self-Compton due to the Klein-Nishina effect \citep{das22, gonzalez22,ren22} but could be explained by proton synchrotron \citep{zhang22_ps}; or photopion decay in the jet \citep{sahu22}; or by ultra-high energy cosmic rays (UHECRs) interacting with  the EBL and CMB photons, and subsequent cascades \citep{das22, alves22}. The intergalactic magnetic field needs to be of the order of $10^{-14}$~G for UHECR protons to be delayed by $\lesssim 2000$~s in order to explain the LHAASO detection. The magnetic field needs to be much lower for UHECR nuclei, and in that case it would require GRB 221009A to have occured in a void with a low intergalactic magnetic field strength \citep{mirabal22}. The universe is expected to be extremely opaque to photons at these energies for the redshift of GRB~221009A, due to $\g\g$ interactions with background radiation fields.  One finds absorption optical depths $\tau_{\rm \g\g}(18\ \TeV)\ga10$ for all recent extragalactic background light (EBL) models \citep[ e.g.,][]{franceschini08, razzaque09, finke10_EBL, kneiske10, dominguez11, helgason12, stecker12, scully14, khaire15, stecker16, franceschini17, andrews18, khaire19, saldana21, finke22}.  These models give a survival probability of $\exp[-\tau_{\g\g}(18\ \TeV)]\la 4.5\times10^{-5}$; the situation is even worse at 251 TeV.

Several ways have been proposed to avoid the $\g$-ray absorption at these energies; one is that the high-energy photons may avoid attenuation by converting to axion-like particles (ALPs) in the presence of magnetic fields in the GRB jet, host galaxy, or intergalactic space \citep{galanti22, zhangma22,baktash22, troitsky22,nakagawa22,carenza22,galanti22_summary}.  Another is through Lorentz invariance violation (LIV), as suggested by \citet{dzhappuev22, baktash22, li22}.

Lorentz invariance is a pillar of special relativity.  It is the principle that there are no preferred inertial reference frames, and physical variables can be transfered from one frame to another with Lorentz transformations.  However, some theories predict LIV, such as supersymmetry, string theory, and other models of quantum gravity \citep[e.g.,][]{Amelino-Camelia1998, amelino01, Mattingly2005, Christiansen2006,Jacobson2006, jacob08, Ellis2008}.
Including LIV, the dispersion relation for photons is modified as
\begin{flalign}
E^2 - p^2c^2 = \pm E^2 \left(\frac{E}{E_{\rm QG}} \right)^n
\end{flalign}
where $E_{\rm QG}$ is an energy, usually thought to be close to the Planck energy, $E_{\rm Planck}=1.2\times10^{28}\ \eV$.  Here $n$ is the order of the leading correction, and the ``$+$'' represents superluminal LIV, and the ``$-$'' represents subluminal LIV \citep[e.g.,][]{martinez20}.  LIV effects are difficult to measure due to the extremely high energies involved; however, nature can produce photons and particles at energies unavailable to terrestrial accelerators, and they propagate through extremely large distances in the universe.  Thus, there are several effects from LIV which are relevant to astrophysics.  One is that the speed of photons becomes energy-dependent.  Time-of-flight measurements from high-energy astrophysical sources have constrained $E_{\rm GQ}$ \citep[e.g.,][]{abdo09_090510,vasil13,ellis19}.  Another relevant effect is the modification of the threshold for the $\g\g$ pair production interaction ($\g + \g \rightarrow e^+ + e^-$).  This modification can decrease the threshold, increasing the absorption optical depth in the superluminal case, and increasing the threshold and decreasing the absorption optical depth in the subluminal case.  Here we are concerned with the subluminal case, which allows the $\g$-ray absorption optical depth $\tau_{\g\g}$ at high energies to be lower than it otherwise would be \citep[e.g.,][]{jacob08}. It is the latter effect that is relevant to the anamalous transparency of very high energy (VHE) photons from GRB~221009A that we explore here.  

In Section \ref{observesection} we present the relevant preliminary observations of GRB~221009A, based on Astronomer's Telegrams (Atels) and Gamma-ray Coordination Network (GCN) circulars.  In Section \ref{Model} we calculate the LIV effect on the $\gamma$-ray flux attenuation and compare with VHE data. We discuss our results and conclude in Section \ref{Discussion}.

\section{Observations}
\label{observesection}

\subsection{{\em Fermi}-LAT}
\label{LATsection}

\begin{figure}
\vspace{2.2mm} 
\epsscale{1.1} 
\plotone{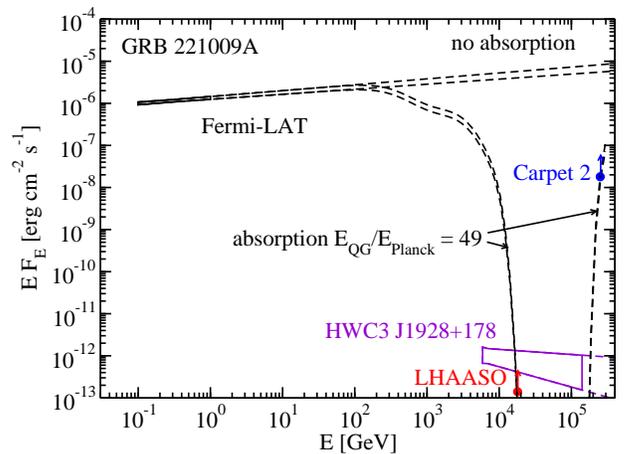}
\caption{The SED for GRB~221009A with the {\em Fermi}-LAT spectrum and the LHAASO, and Carpet 2 95\% lower limits.  The dashed curve indicates the extrapolation of the LAT spectrum to higher energies without $\g\g$ absorption, and with $\g\g$ absorption assuming $E_{\rm QG}/E_{\rm Planck}=49$, as indicated.  The absorption assuming no LIV is identical to this curve at $E<3\times10^5$\ GeV, but does not have the $E>10^5$\ GeV part shown on the plot.
We have also plotted the spectrum for the nearby HAWC source HWC3 J1928+178 and its extrapolation to higher energies.}
\label{sedfig}
\vspace{2.2mm}
\end{figure}

The {\em Fermi}-LAT detected GRB~221009A at 200 -- 800 s after the burst, with a 0.1 -- 1.0 GeV flux of $\Phi_{\rm LAT,tot} = (6.2\pm0.4)\ \times10^{-3}\ \ph\ \cm^{-2}\ \s^{-1}$ and a spectral index of $\G=1.87\pm0.04$ \citep{pillera22}.  The spectrum is described by a power-law, given by
\begin{flalign}
\label{LATspectrum}
\frac{dN}{dE}\biggr|_{\rm LAT} = N_0\left(\frac{E}{E_0}\right)^{-\G}\ ,
\end{flalign}
where we take $E_0=1.0$\ GeV.  The normalization constant $N_0$ can be determined from the integral
\begin{flalign}
\Phi_{\rm LAT,tot} = \int^{E_2}_{E_1}\ dE\ \frac{dN}{dE}\biggr|_{\rm LAT}\ ,
\end{flalign}
where $E_1=0.1$\ GeV and $E_2=1.0$\ GeV.  The 0.1 -- 1 GeV LAT spectrum
for GRB~221009A can be seen in the spectral energy distribution (SED) in Figure \ref{sedfig}.  Since the brightness of this GRB decays with time \citep{ren22,zhang22,zheng22}, this spectrum can be considered an upper limit for the GRB in this energy range at later times.

\subsection{LHAASO}
\label{LHAASOsection}

LHAASO reported the detection of a VHE source within {2000 s of $T_0$ of GRB~221009A, and consistent with its location.  It was detected by both LHAASO's WCDA and KM2A instruments, where the highest energy photon observed by KM2A was $E=18$\ TeV \citep{huang22}.  The effective area of LHAASO-KM2A at 18 TeV is $A_{\rm eff}\approx 0.5\ \km^2$ \citep{lhaaso19}.  Since more photons at these energies may have been detected, we take the implied flux as a lower limit. The Poisson 95\% lower limit for 1 count is $5.13\times10^{-2}$ \citep{gehrels86}.  The observed flux can then be estimated as
\begin{flalign}
\frac{dN}{dE}\biggr|_{\rm obs}(18\ \TeV) & \ga
\frac{5.13\times10^{-2}}{A_{\rm eff} t E} 
\nonumber \\ &
\ga 2.9\times10^{-19}\ \ph\ \cm^{-2}\ \s^{-1}\ \GeV^{-1}\ .
\end{flalign}

Extrapolating the LAT spectrum out to 18 TeV, we find a flux of $9.3\times10^{-12}\ \ph\ \cm^{-2}\ \s^{-1}\ \GeV^{-1}$, much higher than the estimated LHAASO-KM2A flux.  The LHAASO lower limit flux estimate and the LAT extrapolation are plotted in Figure \ref{sedfig}.  
The LAT observation (200-800 s after $T_0$) is not completely overlapping with the LHAASO one (0 to 2000 s after $T_0$).  This is a caveate that should be kept in mind.

\subsection{Carpet 2}

Carpet 2 reported the detection of a 251 TeV photon 4536 s after the GBM trigger, and 1336 s after the {\em Swift}-BAT trigger for GRB~221009A, from a direction consistent with that burst \citep{dzhappuev22}.  The effective area of Carpet 2 depends on source position in the sky; at this energy, the average effective area $A_{\rm eff}=25\ \m^2$ \citep{dzhapp20}.  Using $t=4536\ \s$ and the same procedure as above for LHAASO, for the Carpet 2 detection,
\begin{flalign}
\frac{dN}{dE}\biggr|_{\rm obs} (251\ \TeV)
\ga 1.8\times10^{-16}\ \ph\ \cm^{-2}\ \s^{-1}\ \GeV^{-1}\ .
\end{flalign}
This Carpet 2 lower limit flux estimate is plotted in Figure \ref{sedfig}.
The LAT spectrum, (Section \ref{LATsection}), extrapolated to 251 TeV, is $6.7\times10^{-14}\ \ph\ \cm^{-2}\ \s^{-1}\ \GeV^{-1}$. Since the LAT observation is from an earlier time, and its flux decreases with time, this is a strong upper limit for the flux implied by the 251 TeV photon detected at 4536 s after $T_0$.

\subsection{Nearby HAWC source}
\label{hawcsection}

As reported by \citet{fraija22}, the source HWC3~J1928+178 from the Third HAWC Catalog \citep{albert20_3hwc}, detected up to 140 TeV, is consistent with the reported positions of the LHAASO and Carpet 2 detections.  We plot the spectrum for this source in Figure \ref{sedfig}.  As seen in the figure, the HAWC source is consistent with our estimated LHAASO flux lower limit at 18 TeV, but its extrapolation to 251 TeV is much too faint to be consistent with the lower limit we derived from Carpet 2 detection.  Thus it is unlikely that this source is responsible for the 251 TeV photon detected by Carpet 2.

There is also the possibility that a nearby (presumably Galactic; GRB 221009A had Galactic latitude $b\approx 4.2\arcdeg$) source was flaring contemporaneous with GRB~221009A, and the Carpet 2 detection is from that flare.  But flaring Galactic $\g$-ray sources are rare.  In the Second Fermi All-sky Variability Analysis (2FAVA) Catalog \citep{2fava}, setting aside active galactic nuclei (AGN) and GRBs, there are 73 flares at Galactic latitudes $-10\arcdeg < b < 10\arcdeg$ from known Galactic or unidentified sources in the 7.4 years covered by the 2FAVA catalog. Approximately 1/3 of these flares are from the Crab.  Based on this, the probability of {\em any} Galactic $\g$-ray source flaring at the same time as the Carpet 2 detection is approximately $73 \times (5000\ \s)/(7.4\ \years) \sim 10^{-3}$; and this does not take into account the spatial coincidence.  Thus it is also quite unlikely that the Carpet 2 detection is from a flaring Galactic source.

\section{Gamma-ray absorption and Lorentz Invariance Violation}
\label{Model}

\subsection{Model Calculations}
\label{modelsection}

The $\g\g$ absorption optical depth for $\g$-rays from a source at redshift $z$ with observed dimensionless energy $\e_1=E_1/(m_ec^2)$ with background radiation photons of proper frame energy density $u_p(\e_p;z)$ is given by
\begin{flalign}
\label{taugg_model}
\tau_{\g\g}(\e_1, z) & = \frac{c\pi r_e^2}{\e_1^2 m_ec^2}\ 
\int^z_0 \frac{dz^\prime}{(1+z^\prime)^2}\ 
\left| \frac{dt_*}{dz^\prime}\right|\ 
\nonumber \\ & \times
\int^\infty_{\frac{1}{\e_1(1+z^\prime)}} d\e_{p} 
\frac{ \e_{p}u_{p}(\e_{p};z^\prime)}
{\e_p^{4}}\ 
\bar{\phi}(\e_{p}\e_1(1+z^\prime))\ ,
\end{flalign}
where $r_e\approx 2.82\times10^{13}\ \cm$ is the classical electron radius, $m_e$ is the electron mass, $\bar{\phi}(s_0)$ is a function given by \citet{gould67_crosssec,brown73}, 
\begin{flalign}
\frac{dt}{dz} = \frac{-1}{H_0 (1+z)\sqrt{\Omega_m(1+z)^3 +
    \Omega_\Lambda}}\ ,
\end{flalign}
and we use a flat $\Lambda$CDM cosmology where
$(h,\Omega_m,\Omega_\Lambda) = (0.7, 0.3, 0.7)$, with $H_0 = 100h\ \km\ \s^{-1}\ \Mpc^{-1}$.  Here for $u_p(\ep;z)$ we use the EBL model from \citet{finke22} and the cosmic microwave background (CMB), when appropriate.  Following \citet{jacob08,biteau15}, to include the effects of LIV on $\g\g$ opacity, we allow
\begin{flalign}
\e_1 \rightarrow \frac{\e_1}
{1 + \frac{1}{4}\left(\frac{\e_1 m_e c^2}{E_{\rm QG}}\right)^n\e_1^2}
\end{flalign}
in Equation (\ref{taugg_model}).

\subsection{Results for GRB~221009A}

A common method for constraining EBL absorption is to take the observed spectrum in a region where the EBL is unabsorbed, extrapolate that to a region where it is absorbed, and take that as the highest possible intrinsic flux $dN/dE|_{\rm int}$ \citep[e.g.,][]{chen04, schroedter05_EBL, mazin07, finke09, georgan10, meyer12_EBL, dominguez13_cgrh, abdollahi18, desai19}.  We note that in the 0.1 -- 1.0 GeV energy range, the EBL should be completely transparent to $\g$ rays in all EBL models.  At higher energies, the intrinsic flux is attenuated as
\begin{flalign}
\frac{dN}{dE}\biggr|_{\rm obs} = \frac{dN}{dE}\biggr|_{\rm int}
\exp[-\tau_{\g\g}(E)]\ ,
\end{flalign}
If one has an upper limit on $dN/dE|_{\rm int}$, as described above, then it is possible to constrain the opacity as
\begin{flalign}
\label{taugg}
\tau_{\g\g}(E) < \ln\left(\frac{dN/dE|_{\rm int}}{dN/dE|_{\rm obs}}\right)\ .
\end{flalign}

\begin{figure*}
\vspace{2.2mm} 
\epsscale{1.1} 
\plottwo{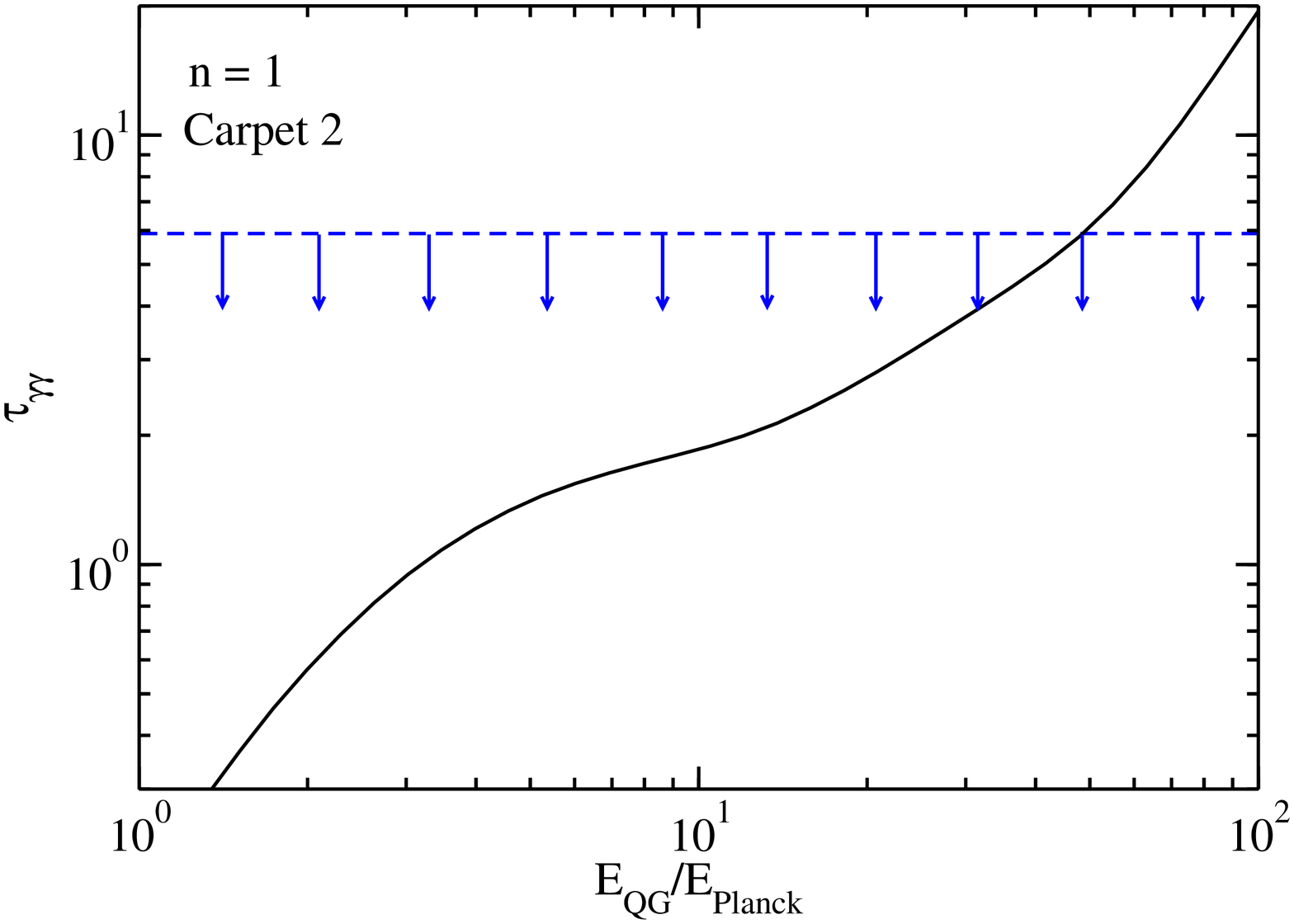}{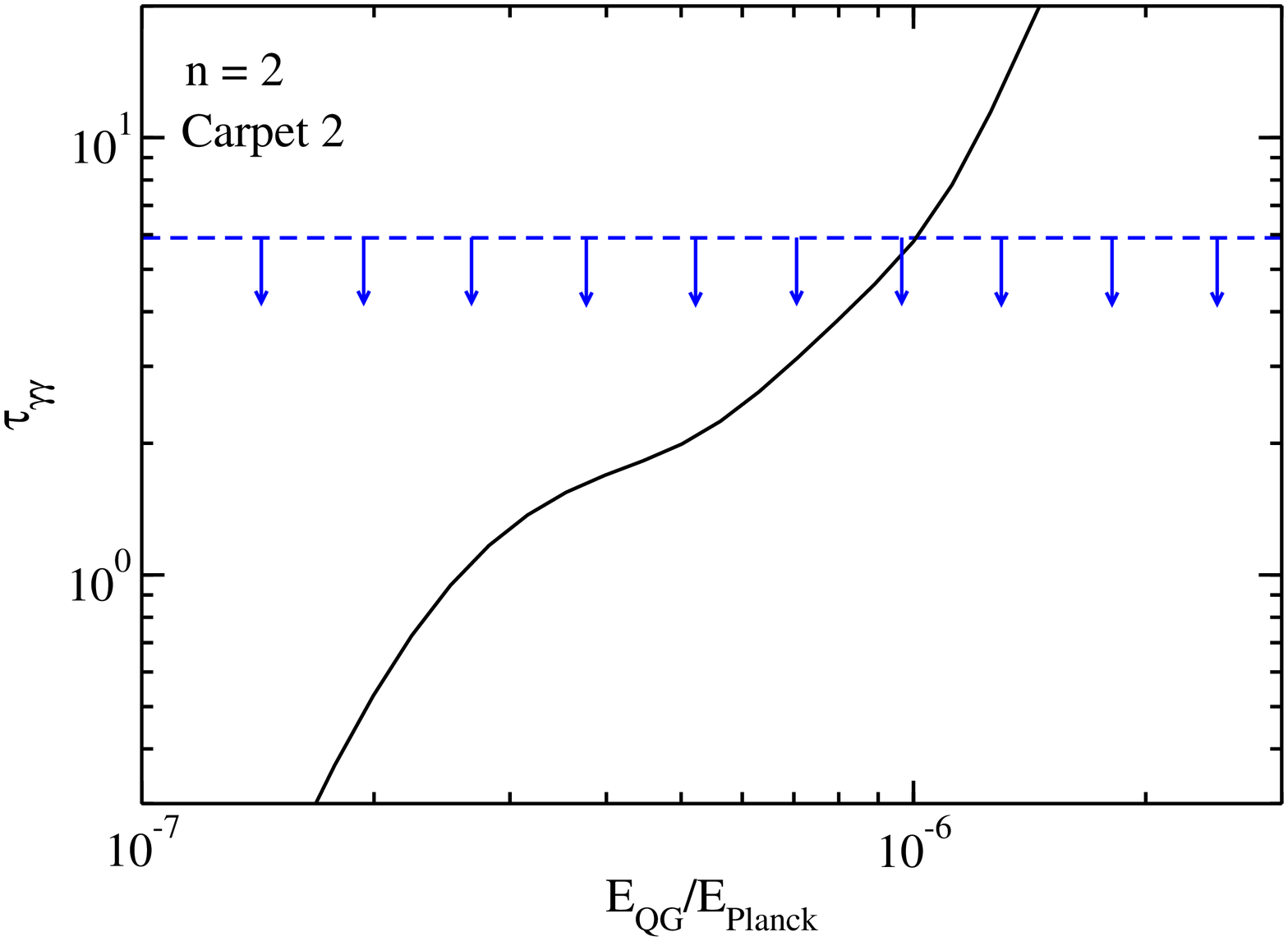}
\caption{The $\g$-ray absorption optical depth at $z=0.15$ and
  $E=251\ \TeV$ (solid black
  curves).  The left figure shows the result assuming the leading order
  correction is linear ($n=1$); the right shows the result assuming
  the leading order correction is quadratic ($n=2$). Dashed blue lines with
  arrows show the $\tau_{\g\g}$ upper limit from the 
  Carpet 2 observations.}
\label{LIVfigure}
\vspace{6.6mm}
\end{figure*}

Using this technique with the LAT spectrum extrapolated to 18 TeV and the LHAASO observation (Section \ref{LHAASOsection}), we get the constraint
\begin{flalign}
\label{taugg_LHAASO}
\tau_{\g\g}(18\ \TeV) \la 17
\end{flalign}
We note that here, and all limits in this paper, are 95\% constraints. For photons at 18 TeV from redshift $z=0.15$, this constraint is consistent with many, but not all, recent EBL models \citep{baktash22} without the need for including LIV.  The infrared EBL relevant here is somewhat uncertain, as reflected in different models.  Our constraint on $\tau_{\g\g}$ here is a bit higher than that of \citet{baktash22}, mainly because we estimate the more conservative 95\% lower limit on flux at 18 TeV. Following the same procedure for the Carpet 2 measurement, we get
\begin{flalign}
\tau_{\g\g}(251\ \TeV) \la 5.9\ .
\end{flalign}

For the 251 TeV photon from redshift $z=0.15$, the relevant photon field for $\g\g$ interactions is the CMB \citep[e.g.,][]{fazio70,protheroe96,dermer09_book}.  Unlike the infrared EBL, the CMB is known to very high precision.  We show the model calculation of $\tau_{\g\g}$ (from Equation [\ref{taugg_model}]) as a function of $E_{\rm QG}/E_{\rm Planck}$ in Figure \ref{LIVfigure} for the Carpet 2 case.  From this figure we can see that the 251 TeV photon from Carpet 2 gives the constraint
\begin{flalign}
E_{\rm QG}/E_{\rm Planck} & \la 49\ (n=1)
\nonumber \\
E_{\rm QG}/E_{\rm Planck} & \la 1.0\times10^{-6}\ (n=2)\ .
\end{flalign}

\section{Discussion}
\label{Discussion}

We have made estimates, based on preliminary observations reported in ATels and GCNs, of $\g$-ray fluxes detected by LHAASO and Carpet 2 from observations of GRB~221009A.  We compared these to the extrapolated LAT spectrum \citep{pillera22} and used these to make estimated constraints on subluminal LIV, particularly on $E_{\rm QG}$.  We use LHAASO and Carpet 2 lower limit flux estimates; if they are significantly larger, the constraints on $E_{\rm QG}/E_{\rm Planck}$ would be lower (and therefore stronger).  Our results do not depend on the detailed spectrum and analysis of the LHAASO and Carpet 2 results, and our assumptions are quite conservative, taking robust 95\% lower limits for the implied flux from the reported photons.  Detailed analysis by the LHAASO and Carpet 2 collaborations will likely strengthen these results, as long as they are not retracted.  Our constraints are broadly consistent with other authors work on constraining $\tau_{\g\g}$ and LIV 
from this burst \citep[e.g.,][]{baktash22,zhao22,galanti22,zheng22}.

If confirmed, this would be the first known upper limit on $E_{\rm QG}$; however, there have been some previous lower limits.  \citet{lang19} found $2\sigma$ lower limits $E_{\rm QG}/E_{\rm Planck}> 10$ ($n=1$) and $E_{\rm QG}/E_{\rm Planck}> 1.9\times10^{-7}$ ($n=2$) using VHE $\g$-ray spectra of blazars detected by imaging atmospheric Cherenkov telescopes.  \citet{vasil13} find $E_{\rm QG}/E_{\rm Planck}>7.6$ ($n=1$) and $E_{\rm QG}/E_{\rm Planck}>10^{-9}$ ($n=2$) from time-of-flight measurements of photons from GRBs. Our results are compatible with all previous $E_{\rm QG}$ lower limits for subluminal LIV. {\em Our result could be the first observational evidence for LIV}.  

However, it does come with a number of caveats.  We assume that
the $\g$-ray spectrum of GRB~221009A is well-behaved at VHEs, 
and that the spectrum does not ``curve up'' above the LAT bandpass; although it is difficult to imagine a GRB being much brighter at these energies.
The results of \citet{lang19} make a similar assumption about the spectra of blazars.  Another possibility is the anomalous transparency could be explained by photon conversion to ALPs, or another mechanism that has yet been proposed.  The Cherenkov Telescope Array (CTA) will be sensitive at $\ga 10$\ TeV and may be able to marginally detect LIV effects in blazar spectra within current LIV constraints, i.e., $10\la E_{\rm QG}\la50$ for $n=1$, especially if the true value is on the lower end of this range \citep{abdalla22}.  It may also be able to confirm or rule out our result with detections of future GRBs, if VHE emission out to 100s of TeV from these sources turns from out to be at all common.

\acknowledgements 

We are grateful to the referees for helpful comments that have improved this manuscript. The authors are grateful to D.\ Alexander Kann for pointing out several minor errors in the version of the manuscript posted on arXiv.  J.D.F.\ would like to thank Elisabetta Bissaldi, Matthew Kerr, Gerald Share, and Jacob Smith for bringing various aspects of GRB~221009A to his attention.  J.D.F.\ was supported by NASA through contract S-15633Y. S.R.\ was supported by a grant from NITheCS and the University of Johannesburg URC. 

\bibliographystyle{apj}
\bibliography{grb_ref,references,ULX_ref,gravwave_ref, mypapers_ref,EBL_ref,liv_ref,blazar_ref,LAT_ref}

\end{document}